\documentstyle[12pt,aaspp4,psfig]{article}
\lefthead{van Breugel et al.}
\righthead{A Radio Galaxy at $z=5.19$ }

\newcommand\tn{TN~J0924$-$2201}

\newcommand\eg{{e.g.,~}}
\newcommand\etal{{et al.~}}

\newcommand\lya{Ly$\alpha$}
\newcommand\CII{\hbox{C~$\rm II$}]~$\lambda$~2326}

\newcommand\Ha{H$\alpha$}

\newcommand\MgII{\hbox{Mg~$\rm II$}~$\lambda$~2800}

\newcommand\OII{[\hbox{O~$\rm II$}]~$\lambda$~3727}
\newcommand\OIII{[\hbox{O~$\rm III$}]~$\lambda$~5007}

\def\deg{\ifmmode {^{\circ}}\else {$^\circ$}\fi}

\def\secper{\ifmmode \rlap.{^{s}}\else $\rlap{.}{^{s}} $\fi}

\def\ergsA{\ifmmode {\rm\,erg\,s^{-1}\,\AA^{-1}}\else
    ${\rm\,erg\,s^{-1}\,\AA^{-1}}$\fi}
\def\ergscmA{\ifmmode {\rm\,erg\,s^{-1}\,cm^{-2}\,\AA^{-1}}\else
    ${\rm\,erg\,s^{-1}\,cm^{-2}\,\AA^{-1}}$\fi}
\def\ergcms{\ifmmode {\rm\,erg\,cm^{-2}\,s^{-1}}\else
    ${\rm\,erg\,cm^{-2}\,s^{-1}}$\fi}
\def\kms{\ifmmode {\rm\,km\,s^{-1}}\else
    ${\rm\,km\,s^{-1}}$\fi}
\def\kmsMpc{\ifmmode {\rm\,km\,s^{-1}\,Mpc^{-1}}\else
    ${\rm\,km\,s^{-1}\,Mpc^{-1}}$\fi}

\def\spose#1{\hbox to 0pt{#1\hss}}
\def\simlt{\mathrel{\spose{\lower 3pt\hbox{$\mathchar"218$}}
     \raise 2.0pt\hbox{$\mathchar"13C$}}}
\def\simgt{\mathrel{\spose{\lower 3pt\hbox{$\mathchar"218$}}
     \raise 2.0pt\hbox{$\mathchar"13E$}}}

\def\oxytwo{[\ion{O}{2}]}

\def\woxytwo{W_{\rm [OII]}^{\rm rest}}

\def\Msun{\ifmmode {\rm\,M_\odot} \else
	${\rm\,M_\odot}$\fi}
\def\msun{\ifmmode {\rm\,M_\odot} \else
    ${\rm\,M_\odot}$\fi}

\begin{document}

\title{A Radio Galaxy at $z=5.19$\altaffilmark{1}}

\author{Wil van Breugel\altaffilmark{2}, Carlos De
Breuck\altaffilmark{2,3}, S.A. Stanford\altaffilmark{2}, \\ Daniel
Stern\altaffilmark{4}, Huub R\"ottgering\altaffilmark{3}, \& George
Miley\altaffilmark{3}}

\altaffiltext{1}{Based on observations at the W.M. Keck Observatory,
which is operated as a scientific partnership among the University of
California, the California Institute of Technology, and the National
Aeronautics and Space Administration.  The Observatory was made
possible by the generous financial support of the W.M. Keck
Foundation.}

\altaffiltext{2}{Institute of Geophysics \& Planetary Physics, Lawrence
Livermore National Laboratory, Livermore, CA 94550, USA; (wil,debreuck,adam)@igpp.llnl.gov}

\altaffiltext{3}{Leiden Observatory, 2300 RA Leiden, The Netherlands; (debreuck,rottgeri,miley)@strw.leidenuniv.nl}
 
\altaffiltext{4}{Department of Astronomy, University of California, Berkeley, CA 94720, USA; dan@bigz.berkeley.edu}
\begin{abstract}

We report the discovery of the most distant known AGN, the radio galaxy \tn\ at
$z = 5.19$.  The radio source was selected from a new sample of ultra--steep
spectrum (USS) sources, has an extreme radio spectral index $\alpha_{\rm 365
MHz}^{\rm 1.4 GHz} = -1.63$, and is identified at near--IR wavelengths with a
very faint, $K = 21.3 \pm 0.3$ object. Spectroscopic observations show a single
emission line at $\lambda \sim 7530$ \AA\, which we identify as \lya.  The
$K$--band image, sampling rest frame $U$--band, shows a multi--component,
radio--aligned morphology, typical of lower--redshift radio galaxies.
\tn\ extends the near--IR Hubble, or $K-z$, relation for powerful
radio galaxies to $z > 5$, and is consistent with models of massive
galaxies forming at even higher redshifts.

\end{abstract}


\section{Introduction}

How did the first objects form after the Big Bang?  In hierarchical
cosmogonies (\eg Turner 1998), the first gravitationally bound systems
may have been stars and small star--forming systems which merge to form
galaxies in large dark matter halos. Arising from the end products of
stellar evolution and mergers, central black holes could grow to become
extremely massive.  However, it is not clear how this process would
work at very high redshifts, where little time is available. It has been
suggested that primordial black holes may form well before their host
galaxies (Loeb 1993).  In any case, accretion events fueling massive
black holes are thought to manifest themselves as active galactic nuclei
(AGN; \eg Rees 1984).  Due to their extreme luminosity, AGN are convenient
beacons for exploring these formative, `Dark Ages' of our Universe.

Extragalactic radio sources have played an important role in identifying
active galaxies at high redshifts. The most distant known {\it galaxies}
have consistently been radio--selected until only very recently. In this
Letter we report the discovery of a radio galaxy at $z = 5.19$.  At this
redshift it is the most distant known AGN, surpassing even quasars for the
first time in 36 years.  Throughout this paper we use $H_0 = 65 h_{65}
\kmsMpc$, $\Omega_M = 0.3$, and $\Lambda = 0$.  For these parameters,
1\arcsec\, subtends 7.0 $h_{65}^{-1}$ kpc at $z = 5.19$ and the Universe
is only 1.08 Gyr old, corresponding to a lookback time 91.1\% of the
age of the Universe.

\section{Source Selection}

The most efficient method to find high--redshift radio galaxies (HzRGs) is to
combine two well--known techniques.  The first is to select radio sources with
ultra--steep spectra (USS) at radio wavelengths, i.e.\ very red radio colors (\eg
Chambers, Miley, \& van Breugel 1990).  Most powerful radio galaxies have radio
spectral energy distributions which steepen with frequency.  Therefore, at fixed
observing frequencies more distant sources exhibit steeper spectra (\eg van
Breugel \etal 1999).

A second selection criterion relies upon the magnitude--redshift
relationship at infrared wavelengths, or $K-z$ Hubble diagram, for
powerful radio galaxies (Figure~\ref{kz}).  At low redshifts ($z < 1$),
powerful radio galaxies are uniquely associated with massive galaxies.
The well--behaved $K-z$ diagram suggests that such galaxies can be found
through near--IR identification.  This has been confirmed by the discovery
of many $3 < z < 4.4$ radio galaxies which approximately follow the $K-z$
relationship, even to the highest redshifts and despite significant
morphological evolution (van Breugel \etal 1998).

Using several new, large radio surveys we constructed a USS sample
($S_\nu \propto \nu^\alpha; \alpha^{\rm 1.4 GHz}_{\rm 365 MHz} <
-1.30$; De Breuck \etal 1999 [DB99]) which is much larger, more accurate,
and reaches fainter flux density limits than previous such samples.
\tn, with $\alpha^{\rm 1.4 GHz}_{\rm 365 MHz} = -1.63 \pm 0.08$, 
is among the steepest sources of our sample.  VLA observations
at 4.85 GHz show the source is a slightly resolved $1\farcs2$ double,
with $S_{4.85GHz} = 8.6\pm0.5$ mJy, centered at $\alpha_{\rm J2000} =
09^h24^m19\fs92$, $\delta_{\rm J2000} = -22\arcdeg 01\arcmin 41\farcs5$
(Figure~\ref{kimage}).

\section{Observations}

We obtained $K_s$ images of \tn\  using NIRC (Matthews \&
Soifer 1994) at the Keck~I telescope. We integrated for 32 minutes
on UT 1998 April 18 in photometric conditions with $0\farcs5$ seeing,
and again for 32 minutes on UT 1998 April 19 through light cirrus with
$0\farcs6$ seeing.  The observing procedures, calibration, and data reduction
techniques were similar to those described in van Breugel \etal (1998).
The final image comprising 3840~s of on--source integration is shown in
Figure~\ref{kimage}. Using circular apertures of $2\farcs1$ diameter,
encompassing the entire object, we measure $K = 21.15$ for night 1,
and 21.45 for night 2. We estimate that $K = 21.3 \pm 0.3$. If \tn\
is at $z = 5.19$ (\S4), then redshifted \OII\ at $\lambda = 2.307\mu$m
would be included in the $K_s$ passband and some of the $K$-band flux 
might be due to line emission.

We obtained spectra of \tn\ through a 1\farcs5 wide, 3\arcmin\ long
slit using LRIS (Oke \etal 1995) at the Keck~II telescope.  The
integration times were 5400~s on UT 1998 December 19 (position angle
0\deg) and 4400~s on UT 1998 December 20 (position angle 180\deg);
both nights were photometric with 0\farcs6 seeing.  The observations
used the 150 lines mm$^{-1}$ grating ($\lambda_{\rm blaze} \approx
7500$ \AA; $\Delta\lambda_{\rm FWHM} \approx 17$ \AA), sampling the
wavelength range 4000 \AA\ to 1$\mu$m.  Between each 1800~s exposure,
we reacquired offset star A (see Fig.~2), performed 20\arcsec\ spatial
shifts to facilitate removal of fringing in the reddest regions of the
spectra, and blind offset the telescope to return \tn\ within the
slit.  We calculated the dispersion using a NeAr lamp spectrum
taken immediately subsequent to the observations (RMS variations of
0.50 \AA), and adjusted the zero point according to telluric emission
lines.  Final wavelength calibration is accurate to 1 \AA.  The
spectra were flux calibrated using observations of Feige~67 and
Feige~110 obtained on each night and were corrected for foreground
Galactic extinction using a reddening of $E_{B - V} = 0.0168$
determined from the dust maps of Schlegel, Finkbeiner, \& Davis
(1998).

We find a strong, single emission line at $\lambda \sim 7530$
\AA\, which shifts by $\approx 16$ \AA\ between the two nights.
(Figure~\ref{spectrum}; Table~1).  The cause of the line offset is
unclear, though it may be related to problems LRIS was experiencing
with slippage in the movable guider at the time of the observations.
The relative brightnesses of other sources on the slit vary between
each 1800~s observation, indicating that despite our precautions of
reacquiring the target after each exposure, guider slippage must have
caused some variations in telescope offsetting.  These slight pointing
changes may have caused the slit to sample different regions of
spatially--extended, line--emitting gas.  Indeed, \tn\ shows two
separate components at $K$ (Figure~\ref{kimage}), and emission--line
regions of HzRGs are known to be kinematically complex (Chambers,
Miley \& van Breugel 1990; van Ojik \etal 1997).

Line parameters are measured with a Gaussian fit to the emission line
and a flat (in $F_\lambda$) fit to the continuum (Table~1).
Equivalent width values were derived from a Monte Carlo analysis
using the measured line flux and continuum values with errors, subject
to the constraint that both are positive. For UT 1998 Dec.\ 19, when no
continuum was reliably detected, we quote the 90\% confidence
limit, $W^{\rm obs} > 2760$ \AA.  For UT 1998 Dec.\ 20, when continuum
was marginally detected, we quote the 90\% confidence interval, $W^{\rm
obs} = 710 - 1550$ \AA. 

\section{Redshift Determination}

As discussed by Dey \etal (1998) and Weymann \etal (1998) for two $z >
5$ \lya-emitting field galaxies, a solitary, faint emission line at
red wavelengths is most likely to be either low-redshift \OII\ or
high-redshift \lya. Similar arguments are even more persuasive for
HzRGs because of their strong, rich emission line spectra. For
example, if the line at $\approx$ 7530 \AA\ were \oxytwo\ at $z =
1.020$ then composite radio galaxy spectra (McCarthy 1993; Stern \etal
1999a) indicate that the \tn\ spectrum should have shown \CII\ at 4699
\AA\, with $\approx 40 - 70$\% the strength of \oxytwo, and \MgII\ at
5653 \AA\, with $\approx 20 - 60$\% the strength of \oxytwo.  Similar
arguments rule out identifying the emission line with \Ha\ at $z =
0.147$ or \OIII\ at $z = 0.504$, since in these cases even stronger
confirming lines should have been seen.

The large equivalent widths also argue against identifying the emission
line with \oxytwo\ at $z = 1.020$, implying $\woxytwo > 1370$ \AA\ (night
1) and $350 < \woxytwo < 770$ \AA\ (night 2).  Radio galaxy composites
typically have rest--frame \oxytwo\ equivalent widths of $\approx 130$
\AA\, (McCarthy 1993; Stern \etal 1999a), though active galaxies with
extreme $\woxytwo$ are occasionally observed ($\woxytwo \approx 750$
\AA; Stern \etal 1999b).  The equivalent width of \tn\ is more typical
of high-redshift \lya\ which is often observed with rest frame values
of several $\times$ 100 \AA\ in HzRGs (Table 2). We also note that the
observations from the second night show that \lya\ is attenuated on the
blue side, presumably due to associated and intervening hydrogen gas,
as is commonly observed in HzRGs (\eg van Ojik \etal\ 1997; Dey 1997)
and normal star-forming galaxies at $z > 5$ (\eg Dey \etal 1998).

Finally, the faint $K$-band magnitude of \tn\ conforms to the
extrapolation of the $K - z$ relation to $z > 5$ (Figure~\ref{kz}).
Identifying the emission line with \oxytwo\ would imply a severely
underluminous HzRG (by 3 -- 4 mag).  Therefore, the most
plausible identification of the emission line in \tn\ is with \lya\ at a
(mean) observed wavelength of 7530 \AA\ and $z = 5.19$. Table~1 gives
the dereddened emission--line fluxes.

\section{Discussion}

Among all known $z \simgt 3.8$ HzRGs, \tn\ is fairly typical in radio
luminosity, equivalent width, and velocity width (Table~2).  But this
source has the steepest radio spectrum, consistent with the $\alpha -
z$ relationship for radio galaxies (\eg\ R\"ottgering \etal 1997).  \tn\
also has the smallest linear size, perhaps indicating that the source is
relatively young and/or embedded in a denser environment compared to the
other HzRGs, commensurate with its large velocity width (van Ojik\etal
1997) and very high redshift.  Together with 8C~1435$+$63, \tn\ appears
underluminous in \lya, which might be caused by absorption in a relatively
dense cold and dusty medium. Evidence for cold gas and dust in some of
the most distant HzRGs has been found from sub--mm continuum and CO--line
observations of 8C~1435$+$63 and 4C~41.17 (\eg Ivison \etal 1998).

\begin{figure}
\figurenum{1}
\psfig{file=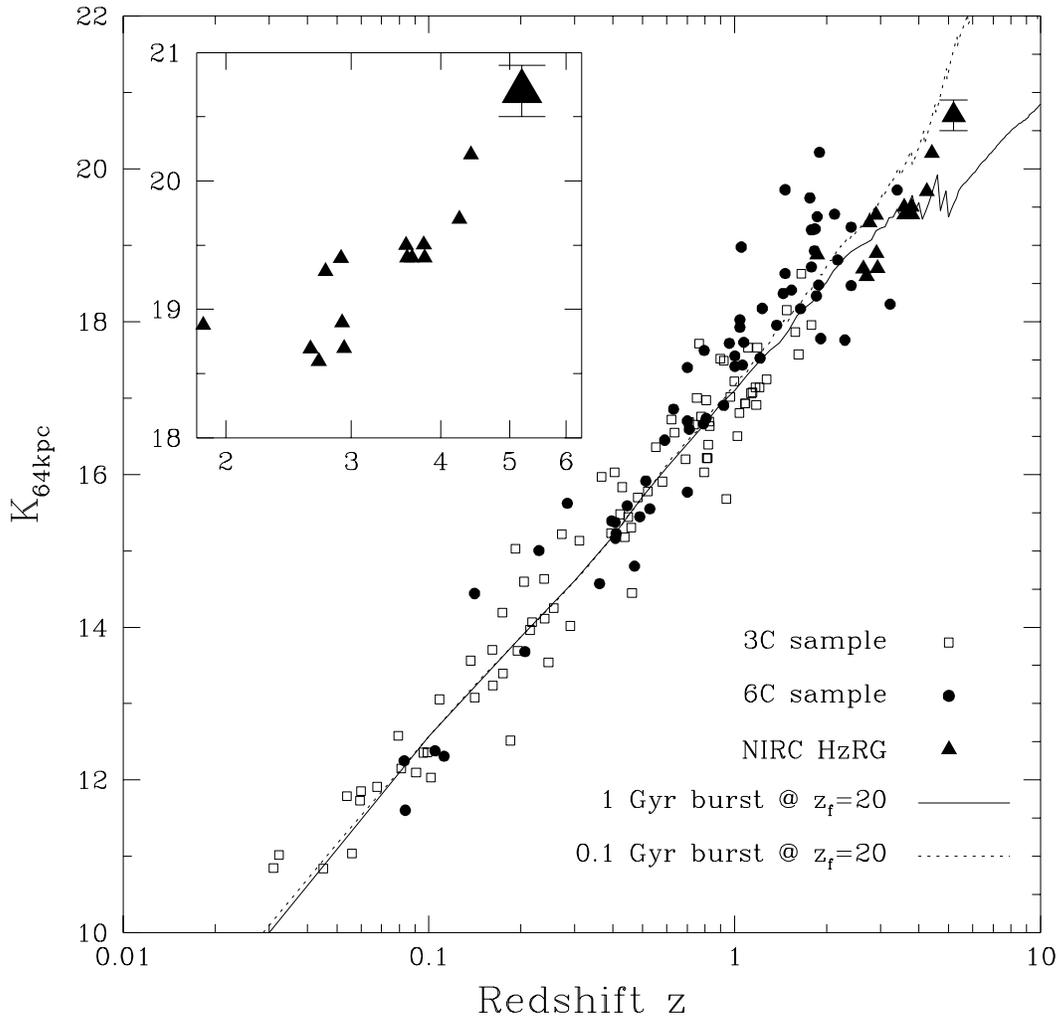,width=16cm}
\caption{Hubble $K-z$ diagram.  Filled triangles are Keck measurements
of HzRGs from van Breugel \etal (1998), the large triangle is \tn,  
and all other photometry is from Eales \etal
(1997). Magnitudes are aperture--corrected to a 64 kpc metric diameter
using $H=65 \kmsMpc$ and $\Omega = 0.30$, for which \tn\ has $K = 20.7 \pm
0.3$.  Two stellar evolution models from Bruzual \& Charlot (1999),
normalized at $z<0.1$, are plotted, assuming parameters as shown.}
\label{kz}
\end{figure}

\begin{figure}
\figurenum{2}
\psfig{file=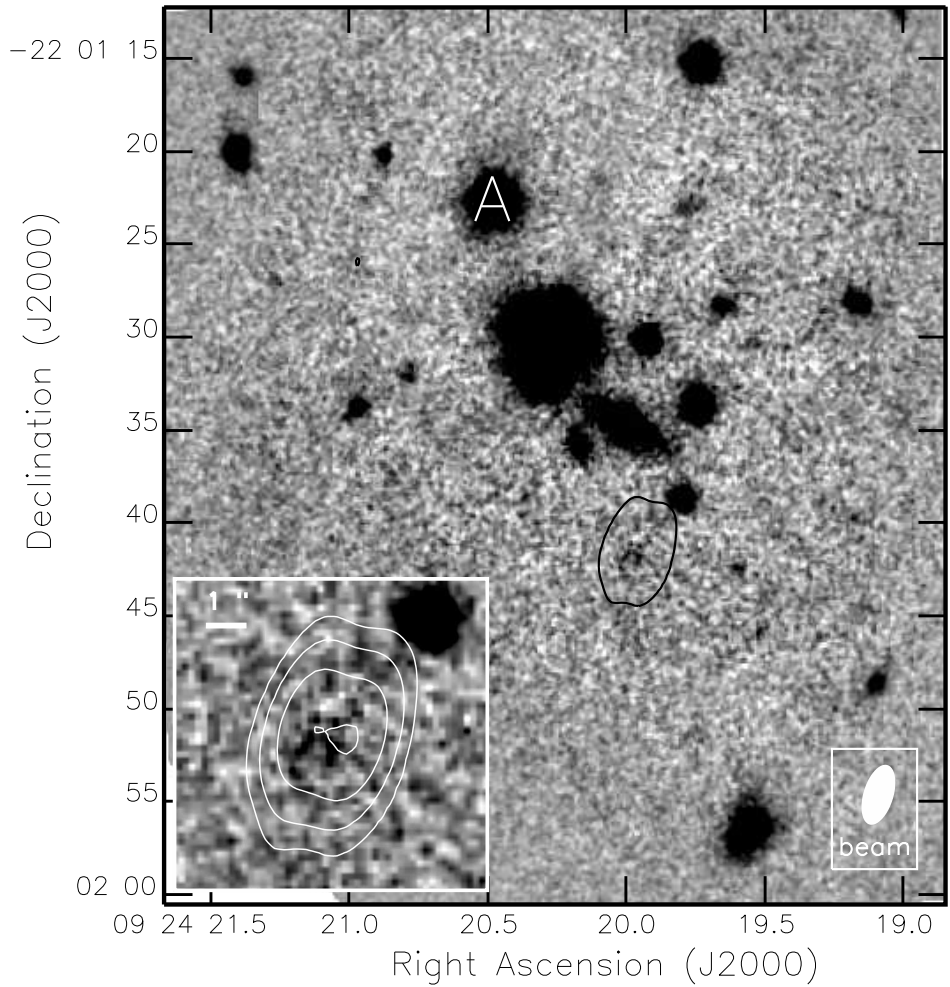,width=16cm}
\caption{Keck/NIRC $K$-band image of \tn, with radio contours superposed.
Star `A', offset from the HzRG
by $\Delta \alpha = 7\farcs2$ (E) and $\Delta \delta = 18\farcs7$ (N),
was used for blind--offsetting for the spectroscopic observations.
}
\label{kimage}
\end{figure}

\begin{figure}
\figurenum{3}
\psfig{file=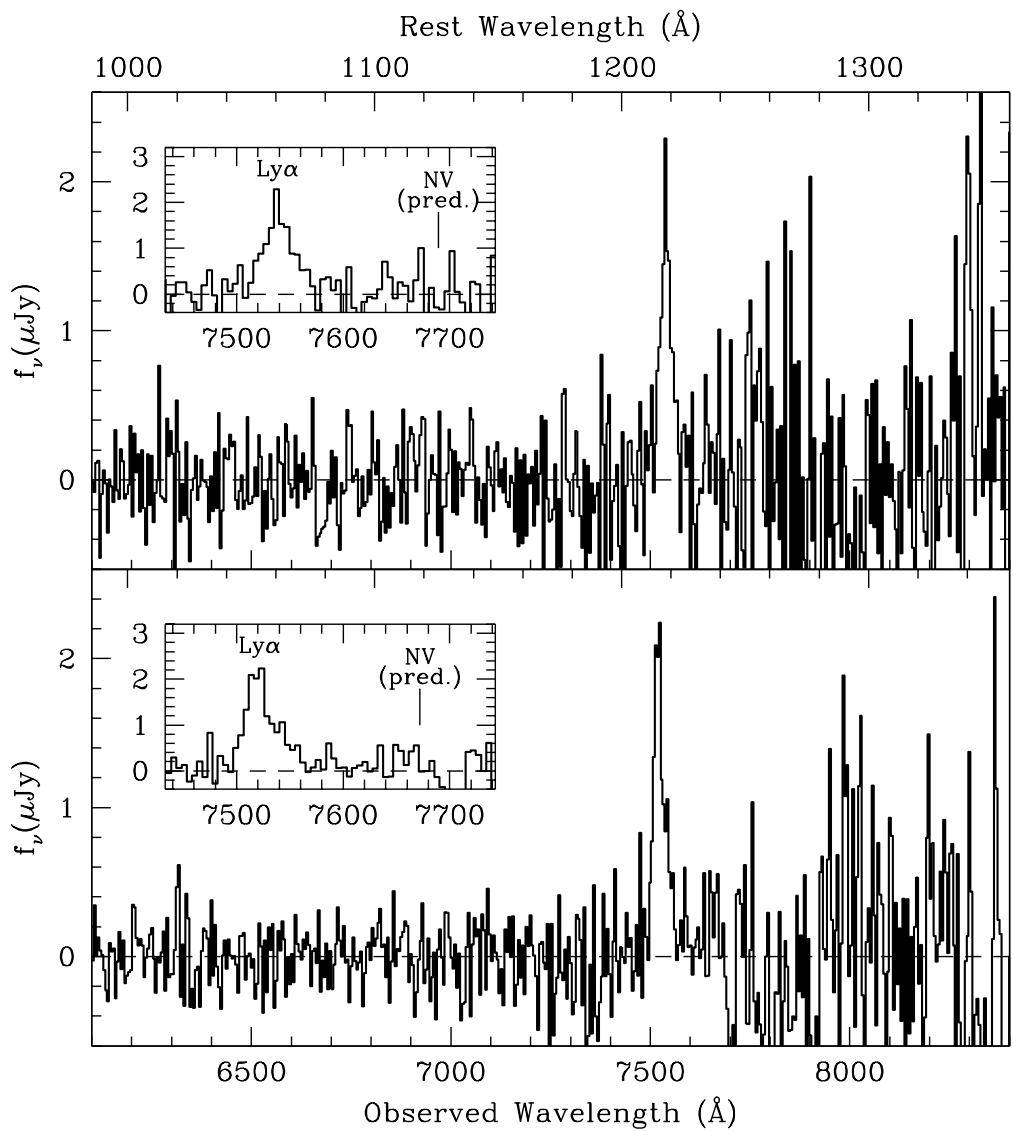,width=16cm}
\caption{Keck spectra of \tn. Top panel is night 1, bottom is night 2. 
Spectra were extracted using a 1\farcs5 $\times$ 1\farcs5 aperture.
Inserts illustrate the emission line morphology, which is attenuated
on the blue edge for night 2.  The predicted wavelength of \ion{N}{5}
$\lambda 1240$ emission is indicated.}
\label{spectrum}
\end{figure}

Our observations of \tn\ extend the Hubble $K-z$ diagram for powerful
radio galaxies to $z = 5.19$.  Simple stellar evolution models are
shown in Figure~\ref{kz} for comparison with the HzRG.  Despite the
enormous $k$--correction effect (from $U_{\rm rest}$ at $z = 5.19$ to
$K_{\rm rest}$ at $z = 0$) and strong morphological evolution (from
radio--aligned to elliptical structures), the $K-z$ diagram remains a
powerful phenomenological tool for finding radio galaxies at extremely
high redshifts. Deviations from the $K-z$ relationship may exist
(Eales \etal 1997; but see McCarthy 1998), and scatter in the $K-z$
values appears to increase with redshift.

The clumpy, radio--aligned $U_{\rm rest}$ morphology resembles that of
other HzRGs (van Breugel \etal 1998; Pentericci \etal 1998).  If the
continuum is dominated by star light, as appears to be the case in the
radio--aligned HzRG 4C~41.17 at $z = 3.798$ (Dey \etal 1997), then
$M(U) = -24.4$ for \tn\.  Then we can derive a SFR of $\sim$200
M$_\odot$ yr$^{-1}$, assuming a Bruzual \& Charlot (1999) GISSEL
stellar evolution model with metallicity $Z = 0.008$, no extinction,
and a Salpeter IMF. This SFR value is highly uncertain due to the
unknown, but competing, effects of extinction and \oxytwo\
emission--line contamination, but is not unreasonable. It is 2.5 times
{\it less} than in 4C~41.17, which has $M(U) = -25.2$ using the same
aperture (Chambers \etal 1990).  \tn\ may be a massive, active galaxy
in its formative stage, in which the SFR is boosted by induced star
formation (\eg Dey \etal 1997).  For comparison other, `normal' star
forming galaxies at $z > 5$ have 10 -- 30 times lower SFR ($\sim 6 -
20 \msun yr^{-1}$; Dey \etal 1998; Weymann \etal 1998; Spinrad \etal
1998).

Recent $z \sim 3$ and $z \sim 4$ Lyman--break galaxy observations have
suggested a possible divergence of star formation and AGN activity at
high redshift (Steidel \etal 1999), contrary to what was previously
thought (\eg Haehnelt, Natarajan \& Rees 1998).  However, if
starbursts and AGN are closely coupled, as suggested to explain the
ultraluminous infrared galaxies (Sanders \& Mirabel 1996), then young
AGN may inhabit especially dusty, obscured galaxy systems.  To obtain
a proper census of the AGN population at the very highest redshifts
therefore requires samples which avoid optical photometric selection
and extinction bias, such as our cm--wavelength/$K$-band radio galaxy
sample.

As emphasized by Loeb (1993), if massive black holes form in a
hierarchical fashion together with their host galaxies, this process
must be quick and efficient, as available timescales are short: at $z
= 5.19$ the Universe is only 1 Gyr old.  It is unclear how this could
be done, so other models, where primordial massive black holes form
soon after the Big Bang and {\it prior} to the beginning of galaxy
formation, may require additional investigation.

\acknowledgments

We thank G.\ Puniwai, W.\ Wack, R.\ Goodrich and R.\ Campbell for their
expert assistance during our observing runs at the W.M.\ Keck Observatory,
and A.\ Dey, J.R.\ Graham and H.\ Spinrad for useful discussions.
The work by W.v.B., C.D.B. and S.A.S.\ at IGPP/LLNL was performed under
the auspices of the US Department of Energy under contract W-7405-ENG-48.
W.v.B.\ also acknowledges support from NASA grant GO 5940, and D.S. from
IGPP/LLNL grant 98--AP017.


\begin{deluxetable}{rrr}
\tablewidth{0pt}
\tablecaption{Spectroscopic Measurements of \tn}
\scriptsize
\tablehead{
\colhead{UT Date} &
\colhead{1998 Dec 19} &
\colhead{1998 Dec 20}}
\startdata
$z\qquad$ $\qquad\qquad\qquad\qquad\qquad\;\;\;~~$ & $5.202 \pm 0.002$ & $5.188 
\pm 0.001$ \nl
$\lambda$~ $\qquad\qquad\qquad\qquad\qquad\;\;\;~~$ [\AA] & $7539.2 \pm 1.9$ & 
$7522.9 \pm 1.5$ \nl
$F_{\rm Ly\alpha}$ $\qquad$ [10$^{-17}$ erg cm$^{-2}$ s$^{-1}$] & $3.4 \pm 0.6$ 
& $3.5 \pm 0.6$ \nl
$F_\lambda^{\rm cont}$ [10$^{-21}$ erg cm$^{-2}$ s$^{-1}$ \AA$^{-1}$] & 
$3.3 \pm 5.7$ & $33.0 \pm 4.9$ \nl
$W_{\rm Ly\alpha}^{\rm obs}$ $\qquad\qquad\qquad\qquad\quad~~~$ [\AA] & $>2760$ & $710 
- 1550$ \nl
FWHM$_{\rm Ly\alpha}$ $\qquad\qquad~~~$ [km s$^{-1}$] & $1574 \pm 192$ & 
$1474 \pm 106$ \nl
\enddata
\label{table}
\end{deluxetable}

\begin{deluxetable}{llccccccl}
\tablewidth{0pc}
\tablecaption{Physical parameters of HzRGs}
\tiny
\tablehead{
\colhead{Name} &
\colhead{$z$} &
\colhead{$L_{\rm Ly\alpha} \times 10^{-43}$} &
\colhead{$L_{\rm 365 MHz} \times 10^{-36}$} &
\colhead{$\alpha_{\rm 365 MHz}^{\rm 1.4 GHz}$} &
\colhead{$W_{\rm Ly\alpha}^{\rm rest}$} &
\colhead{FWHM$_{\rm Ly\alpha}$} &
\colhead{Radio size} &
\colhead{Reference}\\
\colhead{} &
\colhead{} &
\colhead{(erg s$^{-1}$)} &
\colhead{(erg s$^{-1}$)} &
\colhead{} &
\colhead{(\AA)} &
\colhead{(km s$^{-1}$)} &
\colhead{(kpc)} &
\colhead{}\\
}
\startdata
TN~J0924$-$2201 &  5.19 & 1.3 & 7.5 & $-$1.63 &  $>$115 & 1500 &  8 & This paper; DB99 \\
6C~0140$+$326   &  4.41 &  11 & 1.3 & $-$1.15 & \nodata & 1500 & 19 & Rawlings \etal\ (1996) \\
8C~1435$+$63    &  4.25 & 3.2 & 11  & $-$1.31 &     670: & 1800 & 28 & Spinrad \etal\ (1995) \\
TN~J1338$-$1942 &  4.13 &  20 & 2.3 & $-$1.31 &  $>$700: & 1300 & 37 & DB99 \\
4C~41.17        & 3.798 &  12 & 3.3 & $-$1.25 &     100 & 1400 & 99 & Dey \etal\ (1997) \\
4C~60.07        &  3.79 &  16 & 4.1 & $-$1.48 &     150 & 2900: & 65 & R\"ottgering \etal (1997) \\
\enddata
\label{hzrgtable}
\end{deluxetable}

\end{document}